\documentclass[manuscript,screen,nonacm]{acmart}
\AtBeginDocument{%
  }

\setcopyright{none}
\acmDOI{}
\acmISBN{}

\acmDOI{}
\acmISBN{}

\usepackage{tabularx}
\usepackage{eso-pic} 

\usepackage{placeins}
\begin{document}

\title{Medical Knowledge Is Not All You Need: When Medical Q\&A Becomes Situated Patient Assistance}

\author{Shreya Bali}
\email{sbali@andrew.cmu.edu}
\affiliation{%
  \department{Human-Computer Interaction Institute}
  \institution{Carnegie Mellon University}
  \city{Pittsburgh}
  \state{Pennsylvania}
  \country{USA}
}

\author{Riku Arakawa}
\authornote{Both authors contributed equally to this research.}
\affiliation{%
  \department{Human-Computer Interaction Institute}
  \institution{Carnegie Mellon University}
  \city{Pittsburgh}
  \state{Pennsylvania}
  \country{USA}
}

\author{Jill Fain Lehman}
\email{jfl@andrew.cmu.edu}
\authornotemark[1]
\affiliation{%
  \department{Human-Computer Interaction Institute}
  \institution{Carnegie Mellon University}
  \city{Pittsburgh}
  \state{Pennsylvania}
  \country{USA}
}

\author{Alexander K. Maytin}
\email{alexandermaytin@gmail.com}
\affiliation{%
  \institution{University Hospitals of Cleveland}
  \city{Cleveland}
  \state{Ohio}
  \country{USA}
}

\author{Brian Chen}
\email{brianchen2027@u.northwestern.edu}
\affiliation{%
  \department{Computer Science}
  \institution{Northwestern University}
  \city{Evanston}
  \state{Illinois}
  \country{USA}
}

\author{Emma Russell}
\affiliation{%
  \institution{Case Western Reserve University School of Medicine}
  \city{Cleveland}
  \state{Ohio}
  \country{USA}
}

\author{Haarika Reddy}
\affiliation{%
  \institution{Case Western Reserve University}
  \city{Cleveland}
  \state{Ohio}
  \country{USA}
}

\author{Annalise Vaccarello}
\affiliation{%
  \institution{University Hospitals Cleveland Medical Center}
  \city{Cleveland}
  \state{Ohio}
  \country{USA}
}

\author{Dustin P. DeMeo}
\affiliation{%
  \department{School of Medicine}
  \institution{Case Western Reserve University}
  \city{Cleveland}
  \state{Ohio}
  \country{USA}
}

\author{Bryan T. Carroll}
\affiliation{%
  \institution{University Hospitals of Cleveland}
  \city{Cleveland}
  \state{Ohio}
  \country{USA}
}

\author{Mayank Goel}
\affiliation{%
  \department{School of Computer Science}
  \institution{Carnegie Mellon University}
  \city{Pittsburgh}
  \state{Pennsylvania}
  \country{USA}
}

\renewcommand{\shortauthors}{Bali et al.}





\begin{abstract}
Reliability in medical Q\&A is often pursued by grounding responses in authoritative medical information. We show that when Q\&A is embedded within ongoing care, reliability depends on more than what the system knows medically. In a study with 73 skin cancer patients practicing postoperative wound care, 41.9\% of response-requiring questions depended on information beyond the procedure, including visual or physical state, environmental context, or prior actions. These demands varied across patients, consistent with patients recruiting the assistant into different informational roles. We then replayed the questions to seven LLMs while adding procedural and postoperative guidance. Errors remained substantial, including treating unknown states as known, even under explicit guardrails; with full procedural context, six of seven models more often introduced later steps prematurely. Based on these findings, we propose a design space for situated medical assistance that connects what the assistant and patient can each reliably establish to the form of assistance provided. 

\end{abstract}



\maketitle

\section{Introduction}


\begin{figure*}[t]
  \centering
  \includegraphics[width=\textwidth]{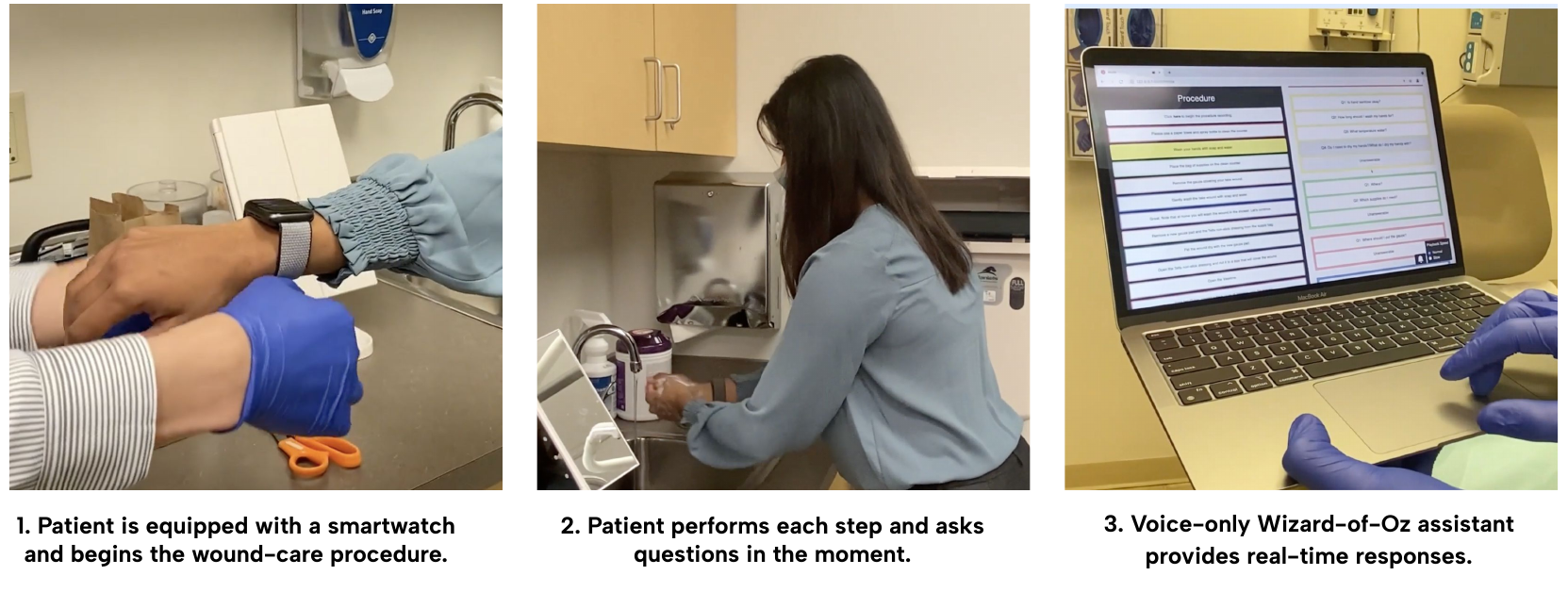}
\caption{Study setup for eliciting situated information needs during postoperative wound-care practice. Patients performed the procedure while wearing a smartwatch, asked questions naturally as they arose during individual steps, and received real-time responses through a voice-only Wizard-of-Oz assistant.}
\Description{
Three photographs illustrate the wound-care study setup. The first shows a
participant wearing a smartwatch while practicing care on a bandage used as a
mock wound. The second shows a participant performing a wound-care step at a clinical workspace. The third shows the researcher-operated laptop interface
used to provide voice-only Wizard-of-Oz assistance. Together, the panels show
that patients performed the procedure while wearing the watch, asked questions
during individual steps, and received real-time voice responses.
}
\label{fig:teaser}
\end{figure*}

Alice has just had surgery and is using a voice assistant to help her manage wound care at home. The assistant has her clinical guidance, her supplies list, and every step of the procedure. It seems unusually well equipped to answer questions about her care. But mid-procedure, Alice holds up an object and asks, \textit{``Is this the tape?''}

Few applications make reliability as non-negotiable as medical Q\&A~\cite{bickmore2018safety,miner2016smartphone}. Accordingly, a central strategy for improving reliability has been to ground models in increasingly complete and trustworthy medical information, such as authoritative clinical guidance, relevant patient context, and task-specific procedural knowledge~\cite{liu2025improving, zakka2024almanac, wu2025clinical}. The complement of that strategy is equally important: a reliable system should decline to answer when the information needed to support an answer is not available~\cite{machcha2026knowing}, though LLMs often fail to abstain~\cite{cocchieri2026llms}. Together, these approaches largely frame reliability around whether the available evidence is sufficient to answer: respond when it is, and refrain when it is not.

In this work, we show that this boundary becomes more complicated when medical Q\&A is embedded in ongoing care. The system must remain useful while avoiding judgments that exceed what the interaction has actually established. Knowing that tape is required at Alice's current step does not establish whether the object she is holding is tape; yet simply refusing because that state is unavailable also fails to provide the assistance she sought. Additional sensing could establish some missing state, but sensed information is not equivalent to authoritative medical evidence: the sensed state may itself be incomplete or uncertain, while richer sensing can introduce privacy tradeoffs~\cite{lau2018alexa}.

We show that this tension recurs frequently when medical Q\&A is embedded in the performance of care. We studied 73 Mohs surgery patients practicing a 14-step wound-care procedure while wearing a smartwatch with a researcher-operated, voice-only Wizard-of-Oz assistant (Figure ~\ref{fig:teaser}) to understand what assistance patients sought while care was being performed and what information those questions required. Across 171 participant-initiated questions, patients sought help not only with what to do, but with identifying objects, recalling prior actions, judging whether actions had been performed adequately, and determining whether they were ready to proceed. Of the 160 questions requiring a response (the rest were deemed to be self-resolved), 41.9\% depended on information about the patient's current state or environment that could not be established from the prescribed procedure alone. Patients also varied in how often their questions required visual-state judgments, suggesting that the informational demands placed on the assistant were shaped not only by the task but by how patients used it in interaction.

We then replayed these questions to seven general-purpose LLMs, progressively adding the complete 14-step procedure and a clinically vetted postoperative-care handout. Mean error fell from 68\% to 55\%, but a decent portion of remaining failures were not simply a smaller version of the same problem. Even the best model erred on 39\% of questions in the richest condition, and adding the procedure made six of seven models more likely to introduce a later step before anything in the interaction established that the current one was complete: 35 responses newly advanced, compared with only 7 that stopped doing so. The procedure therefore made the prescribed trajectory more explicit without making the patient's actual trajectory more observable, redistributing failure even as it reduced overall error. These failures also persisted when we explicitly instructed models to avoid the observed error modes, suggesting that they were not merely a consequence of underspecified prompting.

Together, these findings reframe reliability in situated medical assistance as the problem of remaining useful without acting beyond what the interaction has established. Reliable assistance requires not only authoritative medical information but also sufficient common ground about the patient's unfolding situation. Richer authoritative context reduced errors but could not establish many aspects of the patient's situated state, while explicit guardrails reduced but did not eliminate failures around these unresolved boundaries. Yet simply refusing to answer whenever the needed state is unavailable would leave many patients' questions unresolved. The challenge is therefore to preserve the boundaries of what is known while still providing useful assistance. We argue that situated medical assistance should be evaluated not only by whether its content is medically correct, but by whether its form, certainty, and allocation of epistemic work are warranted by what the interaction has established. Based on our findings, we propose a design space for deciding how unresolved uncertainty should be handled between patient and assistant, considering what each can reliably establish and how that should shape the assistant's response.



Our work contributes:

\begin{enumerate}

\item \textbf{An empirical account of what situated patient assistance requires.}
Across 171 participant-initiated questions from 73 Mohs surgery patients, we distinguish the assistance patients sought from the information needed to provide it. We show that 41.9\% of response-requiring questions depended on situated state beyond the prescribed procedure, and that patients differed in how often they recruited the assistant to make visual-state judgments.

\item \textbf{A characterization of how LLMs fail during situated patient assistance.}
Across seven general-purpose LLMs, we identify recurring failures in correctness, situated grounding, responsiveness, context use, and procedural timing. Adding authoritative procedural and clinical context reduced overall error but left failures around unresolved patient state, while the complete procedure increased premature advancement in six of seven models, revealing that richer context can introduce new failure modes even as it resolves others.

\item \textbf{A reframing and design space for reliability in situated medical assistance.} Based on our empirical findings and expert reference responses, we propose a design space that connects what the assistant and patient can each reliably establish to the form of assistance provided. It illustrates how designers can select among them to support patients without making unsupported judgments about their situated state.

\end{enumerate}
\section{Related Work}

\subsection{Reliability and Grounding in Medical Question Answering}

Medical LLM research has pursued reliability by strengthening the evidence behind a response. Large models now encode substantial clinical knowledge and approach expert-level performance on medical question answering~\cite{singhal2023medpalm,ayers2023comparing}, and conversational systems such as Google's AMIE extend this to multi-turn history-taking, performing on par with physicians in simulated text-based consultations with actors~\cite{tu2025amie}. Grounding responses in retrieved authoritative sources, such as clinical guidelines and vetted literature, further improves factuality and safety~\cite{zakka2024almanac,xiong2024medrag}, and complementary work addresses cases in which available evidence is insufficient through confidence elicitation, calibration, and abstention~\cite{xiong2024confidence,wen2025abstention}. Patient-facing deployments, from public-health check-in calls to voice assistants for older adults, largely inherit this framing~\cite{laranjo2018conversational,fitzpatrick2017woebot,jo2023carecall,yang2024talk2care,pradhan2020use,zubatiy2021empowering}: across medical LLM research more broadly, reliability is often tied to whether generated claims are supported by appropriate medical evidence.

This framing becomes more complicated when medical Q\&A is embedded in the performance of care. Even systems that actively gather information through dialogue, as AMIE does for diagnosis~\cite{tu2025amie}, are evaluated on reaching sound conclusions from what the patient reports; clinical guidance may establish with high confidence what care recommends, while leaving unresolved whether the patient has completed a step, which object they are handling, or whether the current physical state satisfies the conditions under which that recommendation should be acted on. The issue is therefore not simply missing medical knowledge, but that different sources of information establish different aspects of the patient's situation. We examine how this distinction appears in the assistance patients actually seek during postoperative wound care and what happens when increasingly complete medical and procedural knowledge is supplied without resolving the patient's unfolding state.

\subsection{Situated Assistance and Grounding in Ongoing Activities}

Assistants for physical tasks must interpret questions in relation to ongoing activity~\cite{suchman1987plans}: studies of everyday tasks such as guided cooking show that users rely on implicit
conversational cues whose meaning depends on context~\cite{vtyurina2018cues,Jaber2024Cooking,porcheron2018voice}.
Interpreting such questions requires establishing common ground about objects and task state, a collaborative process long studied in human conversation~\cite{clark1991grounding,clark1986referring}. When interpretation fails, users must notice and repair the breakdown themselves~\cite{beneteau2019communication,myers2018patterns}.
Recent systems acquire such context through sensing. GazePointAR uses gaze, pointing, and conversation history to disambiguate references to surrounding objects~\cite{lee2024gazepointar}, while PrISM-Q\&A augments LLM-based response generation with procedural steps inferred from smartwatch audio and motion~\cite{arakawa2024prism}.
These approaches support situated questions without requiring users to fully describe their surroundings or task progress.

More recent assistants combine procedural resources with ongoing observations and dialogue.
Vid2Coach transforms how-to videos into accessible instructions and uses smart-glasses video to answer questions and provide progress feedback~\cite{huh2025vid2coach}, while AROMA combines video recipes, wearable-camera observations, and users' non-visual perceptions to support cooking~\cite{ning2025aroma}.
Follow-up work on PrISM uses mixed-initiative dialogue, including self-narration and confirmation questions, to revise uncertain step estimates~\cite{arakawa2025scaling}.

Building on this work on acquiring and repairing situated context, we examine what remains unresolved even when the procedure and current instructed step are known---and in a domain where unsupported judgments carry clinical consequences. We characterize the naturally occurring assistance patients seek in these moments and show across seven LLMs that richer procedural and clinical information can improve responses without establishing the patient's actual state, and can sometimes lead models to advance before that state has been confirmed.

\section{Study Overview and Research Questions}

We conducted an IRB-approved wound-care training study with 73 patients
undergoing Mohs surgery (27 [37\%] female; mean age 72.5 [9.6 SD]).
While awaiting surgical closure, participants practiced postoperative care on a mock facial wound using the supplies they would later use at home.

\begin{itemize}
    \item \textbf{RQ1:} What assistance do patients seek during physical care,
    and what information does answering it require?
    \item \textbf{RQ2:} What response failures arise when LLMs answer situated patient questions, and how do these failures change as procedural and clinical context is added?
\end{itemize}

\subsection{Study Procedure}

Participants completed the same 14-step wound-care procedure, including cleaning the workspace and wound, preparing supplies, applying Vaseline and a non-stick dressing, securing it with tape, and disposing of materials. A medically trained researcher used a Wizard-of-Oz interface~\cite{dahlback1993wizard} to deliver prerecorded text-to-speech instructions. Participants narrated their actions and requested help as needed; anticipated questions were answered through the interface and unanticipated questions by the researcher. 

Sessions were manually transcribed to remove personally identifiable information and avoid speech-recognition errors. For each question, dialogue context began with the current-step instruction and included the interaction up to the request. We defined a request as either a syntactic question or an utterance that elicited a researcher response; self-resolved questions were retained. This yielded 171 naturally occurring questions from 57 of 73 participants (22 [39\%] female; mean age 72.9 [9.9 SD]).

\subsection{Reference Responses and LLM Evaluation}

For each naturally occurring request, two researchers, including a medical professional, developed an expert reference response appropriate to the information available to the assistant. These post-hoc references were used to characterize appropriate response behavior and evaluate LLM outputs.

We replayed each request under three progressively richer prompt conditions (Table~\ref{table:prompt}), holding the request and dialogue fixed while adding procedural and clinical resources. No condition provided visual access to the participant or environment; situated state was available only when represented in dialogue. Because the task followed a finite 14-step procedure, providing the complete procedure let us distinguish missing procedural knowledge from failures caused by unresolved patient state.

\begin{table}[t]
\caption{Prompt conditions. All included the same interaction instructions, few-shot examples, dialogue context, and patient request.}
\label{table:prompt}
\Description{
The table defines the three progressively richer prompt conditions used in the
LLM evaluation. All conditions contain the same interaction instructions,
dialogue context, and patient request. The second condition additionally
provides the complete 14-step wound-care procedure, and the third additionally
provides a four-page clinically vetted postoperative-care handout. Thus, the
conditions progressively add procedural and clinical information while holding
the interaction itself fixed.
}
\begin{tabular}{p{0.25\linewidth}|p{0.69\linewidth}}
\toprule
\textbf{Condition} & \textbf{Prompt components} \\
\midrule
\textit{Interaction Context}
& Interaction instructions, dialogue context, and request. \\

\textit{Interaction + Full Procedure}
& Interaction condition + the complete 14-step procedure. \\

\textit{Interaction + Full Procedure + Clinical Guidance}
& Interaction condition + Procedure condition + a 4-page clinically vetted postoperative-care handout. \\
\bottomrule
\end{tabular}
\end{table}





\subsection{Analysis}

For RQ1, two researchers iteratively developed a qualitative coding scheme for the form of assistance sought and the information required to answer each question. They independently coded an overlapping subset of 20 questions, compared interpretations, and resolved disagreements through discussion to refine the category definitions. The finalized scheme was then applied across all 171 questions. This process was intended to establish a shared interpretation of the coding scheme rather than to estimate inter-rater reliability. For RQ2, the researchers,  including a medical professional, developed five response-error categories through exhaustive human review of GPT-3.5-turbo outputs ($\kappa=.76$). We then used GPT-4o to apply the finalized categories across seven LLMs~\cite{zheng2023judging}, validating error/no-error judgments on a random 10\% subset on seven models with two human annotators (Fleiss' $\kappa=.67$). As a robustness analysis, we repeated the evaluation on six current models with explicit instructions to avoid the identified error types, using the same questions, context conditions, and evaluation procedure to compare guarded and unguarded responses.

\section{Results}

Our results reveal a mismatch that emerged even in this deliberately bounded care task. We first characterize where these questions exceeded what medical and procedural knowledge alone could establish (RQ1). We then examine how this mismatch was handled in responses and whether progressively richer task and medical context resolved it (RQ2).


\subsection{Patients’ Assistance Needs Extended Beyond the Medical Procedure Information and Varied Across Patients (RQ1)}

\begin{table*}[t]
\caption{We identified two complementary dimensions of patients’ situated questions: the form of assistance sought and the information required to answer. We used these categories to code all 171 naturally occurring questions; information categories were not mutually exclusive.}
\label{tab:assistance-information}
\Description{
The table presents the coding scheme for patient questions in two parts.
Part A defines six forms of assistance sought: procedural execution, object
grounding, task progress, execution trouble, instruction repetition, and
self-resolved requests, with a definition and representative example for each.
Part B defines five types of information that may be required to answer:
current instruction, procedural knowledge, visual or physical state, physical
or environmental context, and prior actions. Information categories are not
mutually exclusive. The scheme distinguishes information obtainable from the
prescribed procedure from situated information about the patient's actual
unfolding task or environment.
}

\small
\setlength{\tabcolsep}{6pt}
\renewcommand{\arraystretch}{1.08}

\begin{tabularx}{\textwidth}{
    @{}
    p{0.23\textwidth}|
    X
    @{}
}

\toprule
\multicolumn{2}{l}{\textbf{A. Form of assistance sought}} \\
\midrule

\textbf{Assistance type} &
\textbf{Definition and representative example} \\
\midrule

Procedural execution &
How to carry out an instructed action or what action to take
(``Does it make any difference what side?''). \\

Object grounding &
Identifying or locating an object referred to during the task
(``Is this the tape?''). \\

Task progress &
Determining whether an action was completed adequately or whether the patient
was ready to proceed
(``Okay, how's that? That's... size?''). \\

Execution trouble &
Overcoming difficulty physically carrying out an action
(``Having trouble getting the backing off the tape.''). \\

Instruction repetition &
Repeating or clarifying the current instruction
(``I did not understand that.''). \\

Self-resolved &
A request the patient resolved without requiring an assistant response. \\

\midrule
\multicolumn{2}{l}{\textbf{B. Information required to answer}} \\
\midrule

\textbf{Information type} &
\textbf{What it establishes} \\
\midrule

Current instruction &
What the patient was instructed to do at the current step. \\

Procedural knowledge &
How an action should be performed, its sequence, or criteria for correct
execution. \\

Visual / physical state &
The current physical condition or result of an action
(e.g., whether a dressing adequately covers the wound). \\

Physical / environmental context &
Objects or features of the immediate environment, including their identity
or location
(e.g., which item is the tape). \\

Prior actions &
What the patient had actually done earlier in the task
(e.g., whether a material had already been used). \\

\bottomrule
\end{tabularx}

\vspace{2pt}
\parbox{\textwidth}{\footnotesize
\textit{Note.} Information categories are not mutually exclusive.
We use \textit{situated information} to refer to evidence about the patient's
actual unfolding task or environment that cannot be established from the
prescribed procedure alone.
}
\end{table*}

\subsubsection{Many Questions Required Situated Information Beyond the Procedure}

Across the 171 questions, patients sought several forms of assistance (Table~\ref{tab:assistance-information}). Qualitatively coding all the questions revealed the following categories: \textit{procedural execution} (85/171), \textit{object grounding} (36/171), \textit{task progress} (9/171), \textit{execution trouble} (6/171), and \textit{instruction repetition} (24/171); a further 11/171 questions were self-resolved. These forms of assistance differed in what information they required. Some questions, such as ``does it make any difference what side,'' could be answered from medical knowledge alone. Others depended on the physical environment or the patient's unfolding state: ``is this the tape''
required identifying an object, while ``okay how's that that's... size'' required both a procedural criterion and the current state of the dressing.

More importantly, knowing the current instruction and procedure was insufficient for 67/160 (41.9\%) questions that required a response. These questions depended on information about the patient's unfolding situation, including visual or physical state (45/160), the physical environment (21/160), or prior actions (5/160). For example, \textit{``Is this the tape?''} required establishing which object the patient was referring to, while \textit{``Okay how's that that's... size''} required both a procedural criterion and the current state of the dressing. Situated information needs therefore cut across assistance types: even procedural questions answerable in principle from the guidance could require evidence about what the patient was holding, had already done, or had successfully completed.




\subsubsection{Patients Appeared to Recruit the Assistant into Different Informational Roles}

Interestingly, patients did not all enlist the assistant in the same way, consistent with the assistant occupying different informational roles across patients. While we did not observe participant-level patterns in the forms of assistance requested, we found differences in the information those questions required. Some patients mostly asked questions answerable from the current instruction or procedure, whereas others more often asked the assistant to assess what was physically happening; for example, whether an object or placement was correct or whether an action had been completed adequately.


Participant-level differences were concentrated in reliance on the assistant for visual-state judgments. As an exploratory analysis, we used step-preserving permutation tests to examine whether these participant-level differences could be explained by the procedural steps at which questions occurred. Among participants with at least two questions, the proportion of visual-state-dependent questions varied significantly across participants, even after accounting for the procedural steps at which questions occurred ($p=.015$). We found no comparable participant-level variation for environmental context ($p=.636$) or prior actions ($p=.842$). Visual-state reliance also persisted across adjacent interactions: when one request required visual or physical-state information, 60.0\% of the participant's subsequent questions did as well, compared with 39.6\% expected under a step-preserving null ($p=.015$). Environmental-context questions likewise clustered locally (28.6\% vs.\ 10.6\%, $p=.032$), despite showing no participant-level differences.

Together, these findings suggest that what an assistant needs to know is shaped not only by where the patient is in the procedure, but by how the patient is using the assistant. Some patients repeatedly recruited it to help establish what was physically happening, effectively expanding its role from providing procedural guidance to making situated judgments. The same task can therefore create different informational demands depending on the role patients assign the assistant in interaction.

\subsection{LLM Responses Often Collapsed Distinctions That Reliable Assistance Preserved (RQ2)}

\begin{table*}[t]
\caption{We identified five recurring ways LLM responses failed during situated patient assistance and used these categories to characterize model responses. Examples contrast an observed LLM response with the expert reference response for the same patient request.}
\label{tab:error-categories}
\Description{
The table defines five LLM error categories for situated patient assistance:
Wrong, Visual, Premature, Unanswered, and Ignored. For each category, the table
provides a definition, explains why the error matters for patient assistance,
and gives an example containing the interaction context, patient request,
observed model response, and expert reference response. The categories capture
incorrect guidance, unsupported claims about unseen state, advancing to later
steps too early, failing to address the patient's immediate question, and
failing to use information already established in the interaction.
}
\small
\setlength{\tabcolsep}{5pt}
\renewcommand{\arraystretch}{1.12}

\begin{tabularx}{\textwidth}{
    @{}
    p{0.085\textwidth}|
    p{0.19\textwidth}|
    p{0.20\textwidth}|
    X
    @{}
}
\toprule
\textbf{Error} &
\textbf{Description} &
\textbf{Why it matters} &
\textbf{Example} \\
\midrule

\textbf{Wrong} &
Provides incorrect, fabricated, contradictory, or potentially harmful medical or procedural guidance. &
Can lead the patient to perform care incorrectly or act on unsupported medical guidance. &
\textbf{Context:} While preparing the Telfa non-stick dressing, the patient was unsure which side should face the wound. \newline
\textbf{Patient:} ``I'm not sure which side is which.'' \newline
\textbf{Claude Haiku 4.5:} ``The non-stick side (which is shiny and slippery) should face directly toward the wound, while the more textured side faces outward.'' \newline
\textbf{Expert:} ``Either side of the non-stick dressing is okay to use.'' \\

\midrule

\textbf{Visual} &
Treats visual or physical state unavailable to the model as established. &
Can turn unresolved situated state into a false premise for subsequent guidance. &
\textbf{Context:} The voice-only assistant had no visual access while the patient was trying to identify the Vaseline packet. \newline
\textbf{Patient:} ``This is Vaseline, right?'' \newline
\textbf{Claude Haiku 4.5:} ``Yes, that appears to be the Vaseline. Please go ahead and open the packet so we can proceed with applying it to the wound.'' \newline
\textbf{Expert:} ``The Vaseline is in a labeled packet.'' \\

\midrule

\textbf{Premature} &
Introduces a later procedural action before the interaction establishes that the current action is complete. &
Can prompt patients to advance too early or require them to retain future instructions while still completing the current step. &
\textbf{Context:} The patient had just removed the gauze and was asking only whether it should be discarded; washing the wound was the subsequent step. \newline
\textbf{Patient:} ``Removing the gauze covering my fake wound---should I throw it away?'' \newline
\textbf{Gemini 1.5 Flash:} ``Yes, you can throw the used gauze away. \textbf{Proceed to gently wash the fake wound with soap and water}.'' \newline
\textbf{Expert:} ``Yes, you can throw the gauze away.'' \\

\midrule

\textbf{Unanswered} &
Fails to address the patient's actual question or assistance need, even when providing otherwise relevant procedural guidance. &
Leaves the patient's immediate uncertainty unresolved and can prevent them from knowing how to act on the instruction. &
\textbf{Context:} The patient needed to pat the wound dry with a new gauze pad but appeared to be confusing the gauze with the Telfa non-stick dressing. \newline
\textbf{Patient:} ``This is the non...'' \newline
\textbf{GPT-4o-mini:} ``Please use the new gauze pad to gently pat the wound dry. Make sure to avoid rubbing, as that can irritate the area.'' \newline
\textbf{Expert:} ``The gauze is a white square of soft fabric. The non-stick dressing is a different item with the word Telfa written on it. Please use the gauze to pat the wound dry.'' \\

\midrule

\textbf{Ignored} &
Treats information already established in the interaction as unavailable or unknown. &
Can contradict prior interaction, trigger unnecessary repetition, or shift avoidable verification work back to the patient. &
\textbf{Context:} The patient had already cleaned the counter using the cleaning spray and a paper towel. \newline
\textbf{Patient:} ``How do you know it's clean?'' \newline
\textbf{GPT-4o:} ``I don't know if the counter is clean. Please ensure the counter is clean by wiping it down...'' \newline
\textbf{Expert:} ``The counter should be clean since you have cleaned it with the cleaning spray and a paper towel.'' \\

\bottomrule
\end{tabularx}
\end{table*}

\begin{figure*}[t]
\centering
\includegraphics[width=\linewidth]{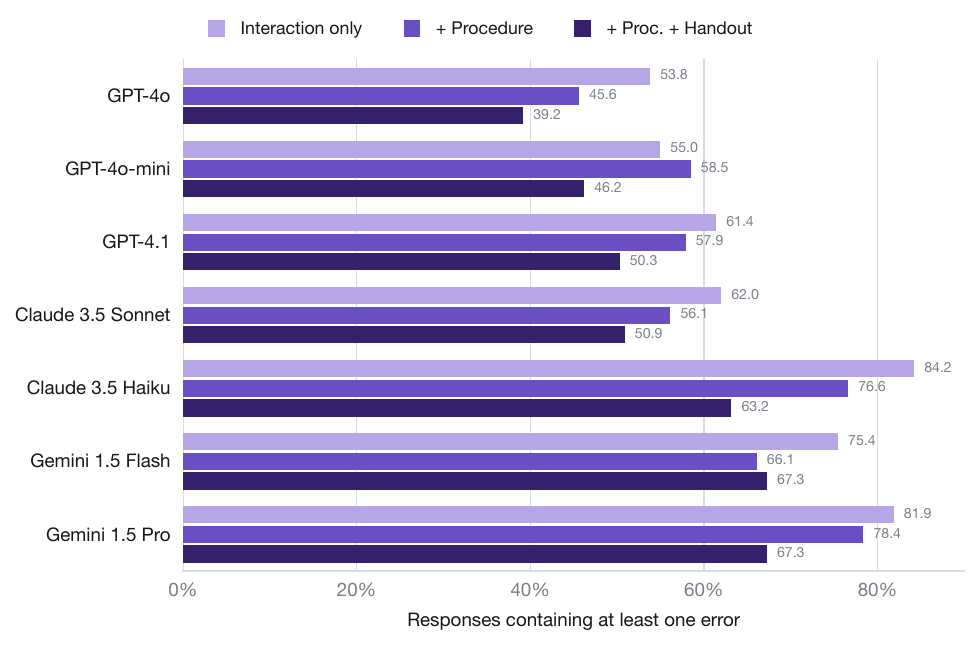}
\caption{Adding medical and procedural context reduced errors across all seven models but did not eliminate them. Bars show the percentage of responses containing at least one error under three progressively richer context conditions.}
\Description{
A grouped horizontal bar chart. Seven LLMs each have three bars, shaded light to
dark for increasingly rich context: interaction context alone, interaction plus
the complete procedure, and interaction plus the procedure and the
postoperative-care handout. The horizontal axis is the percentage of responses
containing at least one error, running from 0 to 90 percent; lower is better.
Models are ordered top to bottom by their error rate in the richest condition.
GPT-4o is best at 53.8, 45.6, and 39.2 percent, followed by GPT-4o-mini at 55.0,
58.5, and 46.2; GPT-4.1 at 61.4, 57.9, and 50.3; Claude 3.5 Sonnet at 62.0, 56.1,
and 50.9; Claude 3.5 Haiku at 84.2, 76.6, and 63.2; Gemini 1.5 Flash at 75.4,
66.1, and 67.3; and Gemini 1.5 Pro at 81.9, 78.4, and 67.3. Every model's darkest
bar is shorter than its lightest one, so all seven improve from interaction
context alone to the richest condition, but two models are not monotonic:
GPT-4o-mini rises from 55.0 to 58.5 percent before falling to 46.2, and Gemini
1.5 Flash falls from 75.4 to 66.1 percent before rising slightly to 67.3. In the
richest condition error rates still range from 39.2 percent for GPT-4o to 67.3
percent for both Gemini models, showing that substantial errors remain even with
the most complete context.
}
\label{fig:model-error-rates}
\end{figure*}


Having established that patient questions could depend on information beyond medical and procedural knowledge alone, we next examined what happened when those questions crossed the boundary of what was available to the responding
system. 


Our evaluation held a single general assistance prompt fixed across conditions
and varied only the task and medical information provided. The results therefore characterize how responses changed as progressively richer context
became available, rather than the best behavior achievable through prompt optimization.


\subsubsection{Expert Responses Preserved Unresolved State While Remaining Useful}

Reference responses curated by a medical expert had access to the same textual information available to the LLMs and no visual context.  We use \emph{unresolved state} to refer to information about the patient's current
situation that is needed to answer a request but has not been established by
the available interaction context. When the information needed to resolve a request was available, they answered directly. When it was
unresolved, they often changed the \emph{form} of the response: providing an identifying description (36/171), a criterion or condition the patient could evaluate (5/171), or asking for clarification when the request itself was ambiguous (4/171).

For example, when a patient asked whether a cut dressing was the right size, the reference response did not confirm an unseen result; it explained that the dressing should completely cover the wound. Similarly, ``Is this the tape?''
was answered with identifying properties rather than confirmation of the unseen object, while whether another Q-tip was needed was conditioned on whether a sufficiently thick layer of Vaseline had already been applied. These responses preserved the unresolved part of the judgment while still helping the patient act on information they could observe.

Under the response prompt used in our evaluation, LLM responses did not always
make the same shift. They often confirmed unseen objects, judged an action complete, or indicated that the patient could proceed despite the relevant state not being established in the dialogue or available context (such as saying `Yes, that is tape'). In such cases, a question whose answer remained underdetermined was rendered as though it had been resolved.

\subsubsection{More Context Reduced Error but Sometimes Changed What Models Got Wrong}

We next examined whether progressively richer task and medical information closed this gap while holding the response prompt fixed. To characterize response failures in this setting, we used five error categories developed through iterative human review for this wound-care context by the study team, including a medical professional (Table~\ref{tab:error-categories}). The categories capture failures in medical correctness, situated grounding, responsiveness, use of available context, and procedural timing; responses exhibiting none of these were considered acceptable. We used GPT-4o to apply five human-developed error categories to 3,591 responses across seven LLMs. The categories were derived from exhaustive coding of GPT-3.5-turbo outputs (Cohen's $\kappa=.76$); GPT-3.5-turbo was excluded from the seven-model comparison because it was used during prompt development. On a random 10\% subset, agreement among two human annotators and the GPT-4o judge on error/no-error was substantial (Fleiss' $\kappa=.67$). Although GPT-4o also appeared among the evaluated models, judge agreement with human consensus was similar for GPT-4o outputs (84.8\%) and the other six models (84.2\%), providing no evidence of systematic self-leniency.

As shown in Figure~\ref{fig:model-error-rates}, all seven models improved in the richest condition, although even the best-performing model still erred on 39\% of responses. Mean error across the seven models fell from 68\% (13\% SD) with \textit{Interaction Context alone}, to 63\% (12\% SD) with \textit{Interaction + Procedure Context}, and to 55\% (11\% SD) after \textit{Interaction + Procedure Context + Postoperative-care handout}. Wrong was the most common category for six of seven models in the richest condition (41–59\% of errors; for Claude 3.5 Sonnet, Wrong and Visual were comparable at 41.4\% and 36.8\%), with Visual (14–37\%) and Premature (7–26\%) accounting for most of the remainder. Medical and procedural incorrectness therefore remained the largest single source of error even with the handout available; situated-state failures (Visual, Premature) accounted for roughly a third of errors. Unanswered and Ignored were rare. Human validation of the judge covered the binary error/no-error distinction only, so these category proportions should be read as descriptive.

The LLM-as-a-judge responses suggested that adding procedural context shifted more answers to the premature category.   We manually reviewed matched responses in which the judge identified a change in premature advancement. After the complete procedure was added, premature advancement was introduced far more often than it was resolved: 35 responses newly introduced a later step, compared with 7 that stopped doing so. This pattern appeared across six of seven models (Gemini 1.5 Pro, GPT-4o, Gemini 1.5 Flash, GPT-4o-mini, Claude 3.5 Sonnet, and GPT-4.1) and was absent only for Claude 3.5 Haiku. Paired examples are provided in Appendix~\ref{app:premature} (Table~\ref{tab:premature-examples}). By contrast, adding the postoperative-care handout did not produce the same shift.

The difference suggests that this was not simply an effect of giving models more information. The procedure told models what came next without establishing whether the patient had finished the current step. For example, when a participant asked \textit{``Do I do it?''} while washing their hands, the interaction-only response stayed focused on hand washing. With the procedure available, the response also introduced the next step, even though nothing in the interaction established that hand washing was complete. The procedure made the prescribed trajectory more explicit without making the patient’s actual trajectory more observable.



Thus, LLMs don't seem to naturally preserve these boundaries, and richer context did not simply remove such uncertainty. 

\subsection{LLM Failures Persisted under Explicit Error Guardrails and Newer Models (RQ2)}

\begin{figure*}[t]
\centering
\includegraphics[width=\linewidth]{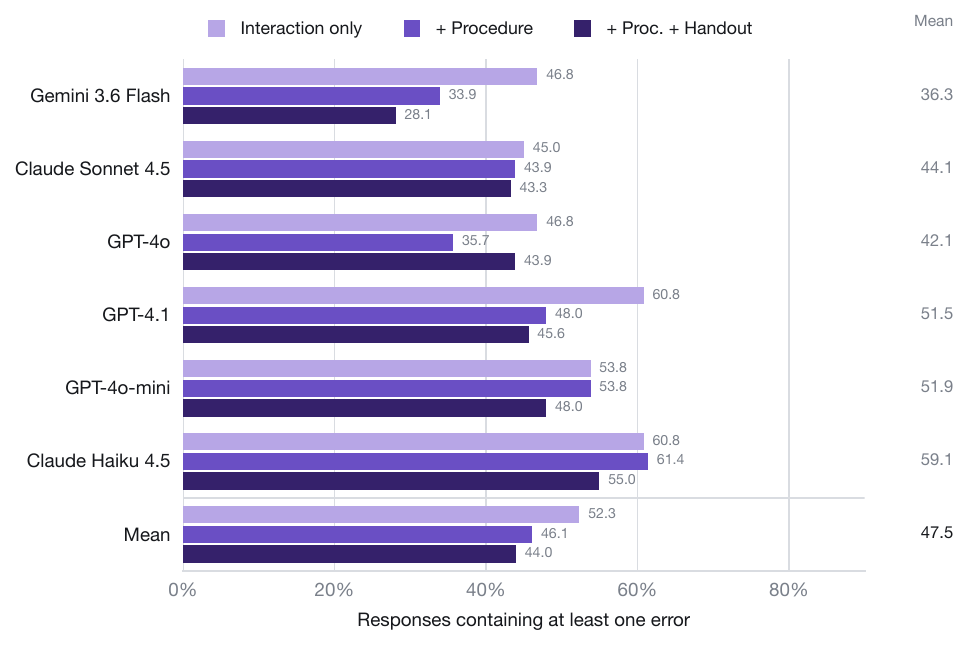}
\caption{Errors persisted across models even with explicit guardrails. Bars show the percentage of responses containing at least one error under each context condition after instructing models to avoid the identified failure modes}
\Description{
A grouped horizontal bar chart with the same layout and axis as the previous
figure, reporting error rates for six LLMs after an explicit guardrail was added
instructing models to avoid the observed failure modes. Each model has three
bars, shaded light to dark for interaction only, plus procedure, and plus
procedure and handout; a column at the right gives each model's mean, and a
ruled-off group at the bottom gives the mean across models. Ordered by the
richest condition, Gemini 3.6 Flash is lowest at 46.8, 33.9, and 28.1 percent
with a mean of 36.3, followed by Claude Sonnet 4.5 at 45.0, 43.9, and 43.3
percent with a mean of 44.1; GPT-4o at 46.8, 35.7, and 43.9 with a mean of 42.1;
GPT-4.1 at 60.8, 48.0, and 45.6 with a mean of 51.5; GPT-4o-mini at 53.8, 53.8,
and 48.0 with a mean of 51.9; and Claude Haiku 4.5 highest at 60.8, 61.4, and
55.0 with a mean of 59.1. The mean error rate across models decreases from 52.3
percent with interaction only to 46.1 percent with the procedure and 44.0 percent
with the procedure plus handout, an overall mean of 47.5 percent. Three models
are not monotonic: GPT-4o drops to 35.7 percent with the procedure but rises to
43.9 with the handout, GPT-4o-mini is unchanged at 53.8 percent across the first
two conditions, and Claude Haiku 4.5 rises from 60.8 to 61.4 percent before
falling to 55.0. Model-level means range from 36.3 percent for Gemini 3.6 Flash
to 59.1 percent for Claude Haiku 4.5, showing that errors remain common despite
the explicit guardrail.
}
\label{fig:llm-as-a-judge-explicit}
\end{figure*}

Our primary evaluation held a general assistance prompt fixed rather than optimizing it against the observed failures. As a robustness check, we repeated the evaluation with an addendum explicitly instructing models not to infer unobserved patient state, to offer descriptions, criteria, or clarification when state was unavailable, and not to advance before the current step was established as complete. We evaluated the three still-available models from our original analysis (GPT-4o, GPT-4.1, GPT-4o-mini) and newer counterparts for the discontinued models (Claude Sonnet 4.5, Claude Haiku 4.5, and Gemini 3.6 Flash), across all three context conditions and 171 questions.

Failures remained common even under these explicit guardrails (Figure~\ref{fig:llm-as-a-judge-explicit}). Mean error across the guarded conditions was 47.5\%. Every model erred on more than a third of questions on average (36.3-59.1\%), and even the best model-context configuration had a 28.1\% error rate. Although the guardrail improved performance, explicitly naming the observed failure modes was insufficient to reliably prevent them.  In fact, examples showing that all five error categories persisted under these
guardrails are provided in Appendix~\ref{app:guardrail-examples}
(Table~\ref{tab:guardrail-examples}).

The remaining challenge is therefore not only instructing models what not to do, but recognizing when the available information does not support the judgment a patient is asking them to make.

Together, these patterns show that LLMs did not reliably preserve the boundary between what the available context established and what remained uncertain. Richer context did not simply remove uncertainty: it resolved some uncertainty, particularly about the prescribed procedure, while leaving the patient’s unfolding state unestablished and enabling new unsupported inferences.

\section{Discussion}

Moving medical AI from answering questions \textit{about} care to assisting patients \textit{during} care changes what a reliable response requires. In our study, patients’ questions often depended on situated state that richer medical and procedural context could not establish. This challenge is likely to become even more pronounced in at-home care, where less controlled environments may introduce additional objects, actions, symptoms, and contingencies that did not arise in our clinic-based practice setting. We therefore discuss implications for how situated medical assistants use context, respond to unresolved state, and should be evaluated.

\subsection{Reliability Changes Character When Medical Q\&A Becomes Embedded in Care}

\begin{figure*}[t]
  \centering
  \includegraphics[width=\textwidth]{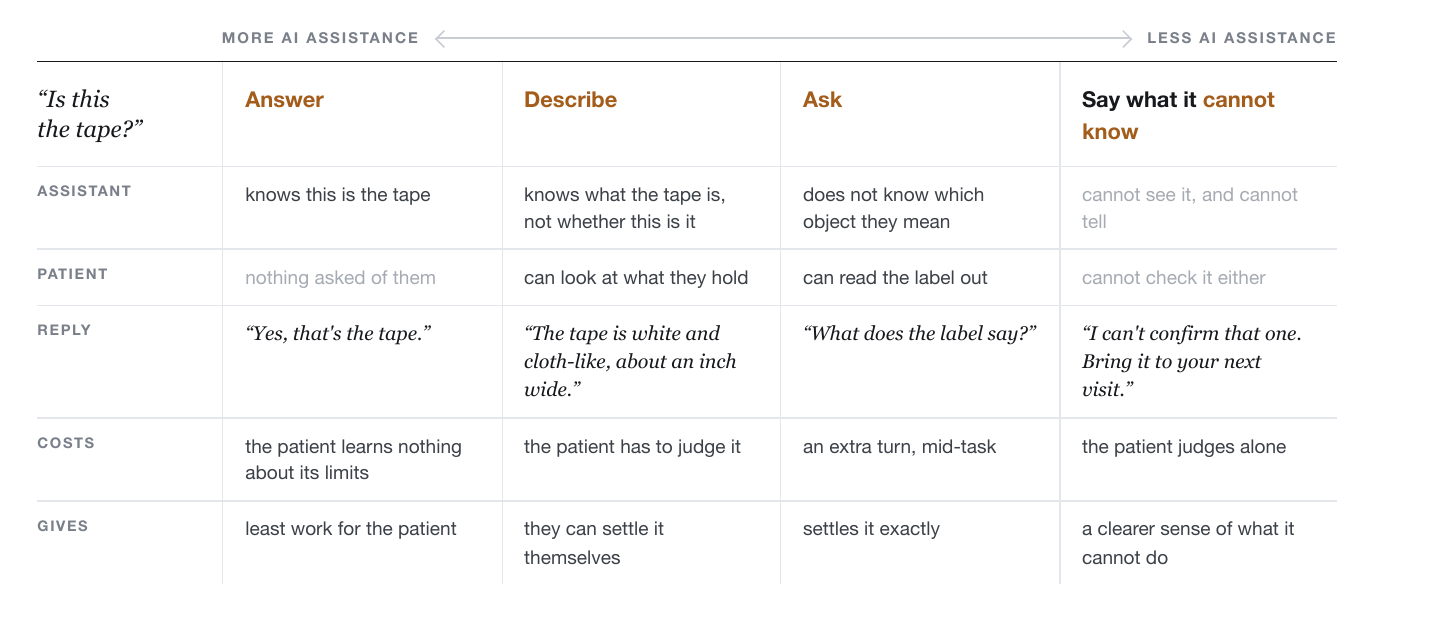}
\caption{\textbf{Proposed design space for allocating judgment between patient and assistant in situated medical assistance.} The appropriate response depends on both what the assistant can establish and what the individual patient can reasonably determine. The design space illustrates how assistance can range from the assistant providing a direct answer to supporting the patient's own judgment, seeking clarification, or acknowledging when neither can reliably resolve the uncertainty.}
\Description{
A design-space matrix uses the example question ``Is this the tape?'' to show
four response strategies arranged from more to less assistant initiative:
Answer, Describe, Ask, and Say what it cannot know. For each strategy, rows
specify what the assistant knows, what the patient must determine, an example
reply, the cost to the patient, and the benefit. Answer requires the assistant
to know that the object is tape and minimizes patient work. Describe is
appropriate when the assistant knows what the tape should look like but cannot
see the object, allowing the patient to compare identifying features. Ask is
appropriate when the assistant needs additional information, such as the
object's label. Saying what it cannot know is used when neither party can
reliably establish the needed state. The strategies illustrate a tradeoff
between assistant initiative and epistemic certainty.
}
\label{fig:design-space}
\end{figure*}

\begin{figure*}[t]
  \centering
  \includegraphics[width=\textwidth]{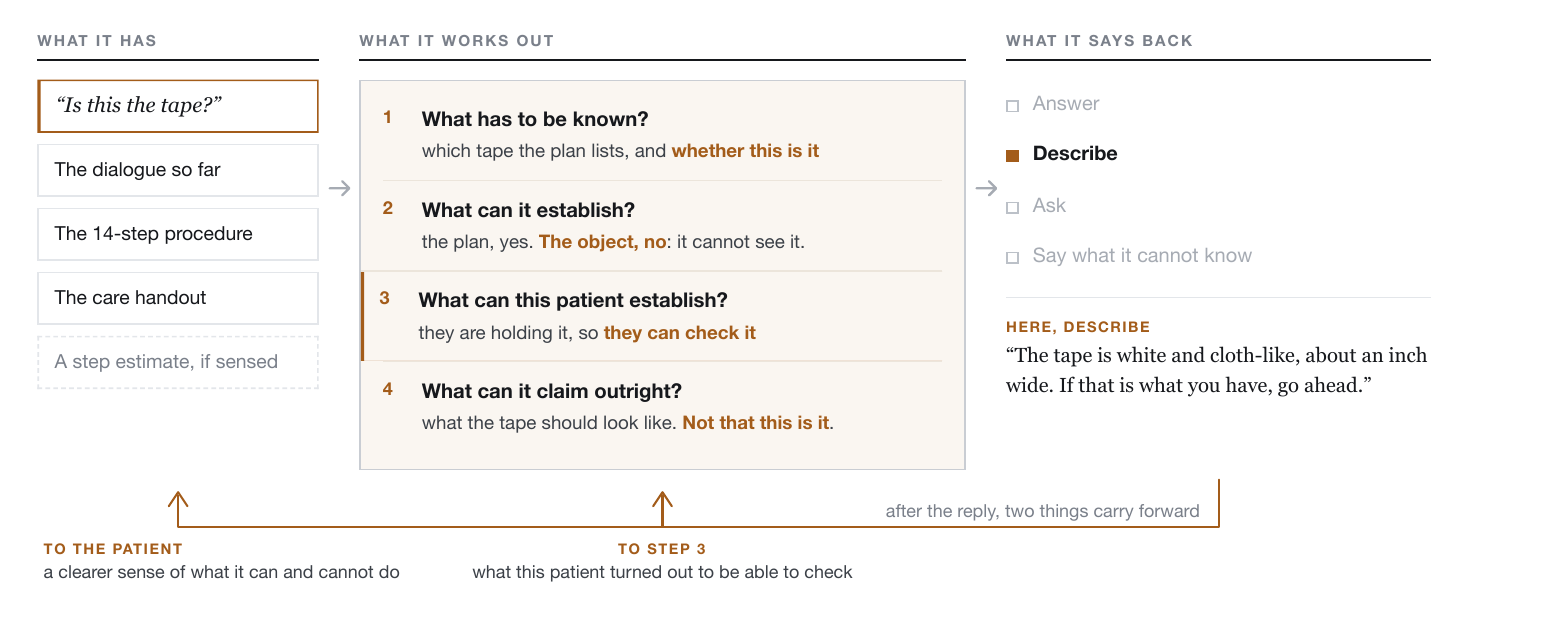}
\caption{Illustrative example of applying epistemic allocation to a situated patient request. Applying the design space to “Is this the tape?” The assistant separates what the medical resources establish from what remains ungrounded in the interaction, considers whether the patient can resolve the remaining uncertainty, and selects a response form accordingly. What the patient successfully establishes can inform subsequent assistance.}
\label{fig:worked-example}
\Description{
A worked example applies the proposed epistemic-allocation process to the
question ``Is this the tape?'' The assistant has the dialogue so far, the
14-step procedure, the care handout, and potentially a sensed step estimate.
It reasons that answering requires knowing both which tape the care plan
specifies and whether the object the patient is holding is that tape. It can
establish the care-plan information but cannot see the object, while the
patient can inspect the object. The assistant therefore selects the Describe
strategy rather than directly confirming the object and replies that the tape
is white, cloth-like, and about an inch wide, asking the patient to compare
those properties. The interaction can then carry forward both what this
patient can successfully check and a clearer understanding of the assistant's
perceptual limits.
}
\end{figure*}

Medical AI is often designed toward reliability by grounding responses in authoritative clinical information. Our findings show a tension in extending this strategy to situated care. An assistant may have highly reliable information about what care is appropriate while remaining uncertain about whether the patient has completed a step, which object they are holding, or
whether an action was performed adequately. These are not simply different amounts of context: they are different kinds of knowledge with different
grounds for certainty. 


This creates a particular tension for patient-facing medical AI. Being too willing to infer across these sources risks encouraging inappropriate reliance on judgments that are not supported by the patient’s current situation ~\cite{lee2004trust,bucinca2021trust}. Being too conservative, however, could lead the assistant to repeatedly defer, clarify, or ask the patient to verify state, shifting substantial work back to someone already performing an unfamiliar care task. The challenge is therefore
not simply to maximize certainty, but to determine which claims can be made reliably from which sources, and when remaining uncertainty should change the form or initiative of the response~\cite{horvitz1999mixed,kay2016ish}. 

This tension may also be partly a product of how such assistants are embodied. Our setup placed the assistant on the body, co-present with the task. Participants were given no account of its perceptual capabilities. We cannot attribute the prevalence of situated requests to this configuration, since we did not compare it against a stationary or screen-based assistant. But the pattern suggests a hypothesis worth testing: an assistant that accompanies the patient through the performance of care may be treated as a co-present participant with perceptual access, whereas one consulted between episodes of care is more likely to be treated as a resource for information about care. If so, the form factor that makes situated assistance useful is also what generates demands it cannot meet.

\subsection{Unresolved State Should Change the Form of Assistance
}


Prior work on situated task assistance has largely focused on acquiring more of the context needed to answer a request, for example by inferring task
progress from sensing or using mixed-initiative dialogue to repair uncertain state estimates~\cite{arakawa2024prism, arakawa2025scaling}. These approaches are important for providing assistants with as much relevant information as possible. But what should an assistant do when the state needed to answer a request has not been established reliably? This question matters both when additional sensing is undesirable, for example, because of privacy or acceptability concerns and when available sensing cannot establish the relevant state with sufficient reliability (due to accuracy problems).

Our findings highlight a complementary solution as well: even when additional context cannot be obtained reliably, an assistant may still be able to help by changing the \emph{form} of assistance it provides. The expert reference responses did not simply answer or abstain when relevant state was unavailable. They instead
provided identifying descriptions, exposed criteria that patients could evaluate themselves, conditioned guidance on information patients could observe or report, or requested clarification when the request itself was
ambiguous. For example, rather than confirming that an unseen dressing was the correct size, the response could explain that it should fully cover the wound; rather than claiming that an unseen object was tape, it could describe the properties by which the patient could identify it. In these cases, the assistant preserved the unresolved part of the judgment while transforming
the request into one the patient might be able to resolve.  

However, whether such a transformation is actually useful depends on what the patient can reasonably establish. A criterion that is sufficient for one patient may be difficult for another to interpret, perceive, or apply while performing an unfamiliar care task. This suggests that response selection should depend not only on \emph{what information is missing}, but also on \emph{who can reliably resolve it}. Rather than adopting a fixed policy of answering, asking, or abstaining, situated assistants could allocate the
remaining epistemic work between system and patient depending on both the questions and the patient: answering directly when the relevant state is established; exposing a description or criterion when the patient can plausibly evaluate it; asking a targeted question when a
small amount of patient-provided information could resolve the uncertainty; and withholding, seeking additional sensing, or escalating when neither party can establish the needed state reliably. 

Importantly, this allocation need not remain fixed. An assistant could update its estimate of what a particular patient can resolve based on whether previous descriptions, criteria, or clarification questions were successfully acted upon, rather than repeatedly placing the same verification burden on the patient. Our study does not test such adaptation directly, but the variation in situated information needs and the response strategies observed in the expert references motivate it as a design direction. More broadly, this reframes situated assistance as an \emph{adaptive allocation of epistemic work}: reliable assistance requires reasoning not only about what remains unknown, but about who can reliably resolve that uncertainty and how the interaction should change when they
cannot. Figure~\ref{fig:design-space} summarizes this design space, showing how
the form of assistance can vary depending on what the assistant and patient
can each reliably establish, and Figure~\ref{fig:worked-example} illustrates this process for the recurring
request ``Is this the tape?''

\subsection{Situated Assistance Makes Known LLM Failure Modes Routine}

Our case study connects situated medical assistance to longstanding concerns about whether AI systems can recognize and communicate the boundaries of their knowledge and capabilities~\cite{amershi,yang2020reexamining,kocielnik2019imperfect}. Many failures we observe similarly reflect known LLM limitations, including unsupported claims~\cite{10.1145/3703155}, failure to abstain~\cite{cocchieri2026llms}, and poor uncertainty calibration~\cite{xiong2024confidence}. Our findings show how situated assistance can turn these limitations into routine interaction problems.

Retrieval and grounding can supply factual and procedural knowledge~\cite{liu2025improving,zakka2024almanac,wu2025clinical}, but patients may shift from asking what to do to asking \textit{Is this the tape?} or \textit{Did I do that right?} Such questions depend on surroundings, prior actions, or physical state that may not be available in any retrievable source. The challenge is therefore not only whether the model knows the relevant medical information, but whether it recognizes when that information is insufficient to support the judgment being requested.

Situated care makes this boundary consequential and recurrent. Appropriate and inappropriate responses may contain similar factual content while differing in whether they accept an unsupported premise, preserve unresolved uncertainty, or take unwarranted initiative. Reliable situated assistance therefore requires evaluating not only factual correctness, but whether a response is \emph{warranted by what the system can actually establish}.

\subsection{Limitations and Future Work}

Our study examines a single, deliberately bounded postoperative-care task
performed on a mock wound with Wizard-of-Oz assistance while participants wore a watch. This design allowed us to observe naturally occurring questions while holding the procedure
relatively constant, but it does not capture the full complexity or risk of performing wound care at home. Participants were predominantly older adults recruited from one clinical setting, and interactions occurred in English;
assistance needs may differ across populations, care tasks, and settings.

The practice setting likely also shaped which assistance needs arose. Because participants worked on a mock wound in clinic while awaiting true wound closure, our taxonomy primarily captures procedural, object-grounding, and task-state questions that arise during learning and performing wound care. At home, patients may additionally ask about pain, bleeding, infection, healing, anxiety, or whether clinical escalation is needed. We therefore do not treat the categories observed here as an exhaustive taxonomy of postoperative assistance needs.

Our LLM evaluation replayed these interactions textually and therefore did not test a deployed multimodal assistant with cameras, wearables, or other sensing.
Such systems may resolve some of the uncertainties observed here, while
introducing uncertainty of their own through missing or noisy observations.
Finally, the reference answers reflect expert guidance developed within one
clinical context. Future work should examine agreement across clinicians and
institutions, particularly for questions where more than one response strategy
may be clinically appropriate.

\section{Conclusion}
When medical Q\&A is embedded within ongoing care, patients may ask an assistant to do more than provide medical or procedural information. In our study, patients sought help identifying objects, recalling prior actions, assessing task progress, and interpreting physical state, creating a mismatch between the role they recruited the assistant to play and what its available information could establish. Our LLM evaluation showed the consequences of this mismatch: additional procedural and clinical knowledge improved many responses, but could not resolve missing situated state and sometimes encouraged action beyond what the interaction supported. We therefore reframe reliability in situated medical assistance around whether a response remains warranted by what has actually been established. Building on this reframing, we propose a design space for allocating unresolved epistemic work between patient and assistant, enabling systems to remain useful by answering when a judgment is grounded, shifting to descriptions or criteria when patients can resolve what remains unknown, seeking targeted information when additional grounding is possible, and withholding or escalating when neither party can reliably establish the needed state.

\section{AI Usage Disclosure}
Generative AI tools were used in the preparation of this work. ChatGPT (OpenAI) and Claude (Anthropic) were utilized throughout sections of the paper to improve writing, spelling, grammar, and the phrasing, organization, and flow of information, including transforming detailed outlines and notes into paragraph form. Claude Code (Anthropic) was used to assist with the development of code and analysis scripts used in this work. Any generated content or code used was reviewed and further modified by the authors to accurately represent the work, ensure originality and correctness, and align with the authors’ writing style and intended analyses.


\bibliographystyle{ACM-Reference-Format}
\bibliography{main}

@inproceedings{amershi,
author = {Amershi, Saleema and Weld, Dan and Vorvoreanu, Mihaela and Fourney, Adam and Nushi, Besmira and Collisson, Penny and Suh, Jina and Iqbal, Shamsi and Bennett, Paul N. and Inkpen, Kori and Teevan, Jaime and Kikin-Gil, Ruth and Horvitz, Eric},
title = {Guidelines for Human-AI Interaction},
year = {2019},
isbn = {9781450359702},
publisher = {Association for Computing Machinery},
address = {New York, NY, USA},
url = {https://doi.org/10.1145/3290605.3300233},
doi = {10.1145/3290605.3300233},
booktitle = {Proceedings of the 2019 CHI Conference on Human Factors in Computing Systems},
pages = {1–13},
numpages = {13},
location = {Glasgow, Scotland Uk},
series = {CHI '19}
}

@article{10.1145/3703155,
author = {Huang, Lei and Yu, Weijiang and Ma, Weitao and Zhong, Weihong and Feng, Zhangyin and Wang, Haotian and Chen, Qianglong and Peng, Weihua and Feng, Xiaocheng and Qin, Bing and Liu, Ting},
title = {A Survey on Hallucination in Large Language Models: Principles, Taxonomy, Challenges, and Open Questions},
year = {2025},
issue_date = {March 2025},
publisher = {Association for Computing Machinery},
address = {New York, NY, USA},
volume = {43},
number = {2},
issn = {1046-8188},
url = {https://doi.org/10.1145/3703155},
doi = {10.1145/3703155},
journal = {ACM Trans. Inf. Syst.},
month = jan,
articleno = {42},
numpages = {55}
}

@inproceedings{cocchieri2026llms,
  title={LLMs (Almost) Never Abstain Under Medical Uncertainty},
  author={Cocchieri, Alessio and Ragazzi, Luca and Tagliavini, Giuseppe and Moro, Gianluca},
  booktitle={Proceedings of the 64th Annual Meeting of the Association for Computational Linguistics (Volume 1: Long Papers)},
  pages={29573--29613},
  year={2026}
}

@inproceedings{machcha2026knowing,
  title={Knowing when to abstain: medical LLMs under clinical uncertainty},
  author={Machcha, Sravanthi and Yerra, Sushrita and Gupta, Sahil and Sahoo, Aishwarya and Sultana, Sharmin and Yu, Hong and Yao, Zonghai},
  booktitle={Proceedings of the 19th Conference of the European Chapter of the Association for Computational Linguistics (Volume 1: Long Papers)},
  pages={6153--6182},
  year={2026}
}

@article{wu2025clinical,
  title={Clinical pathway-aware large language models for reliable and transparent medical dialogue},
  author={Wu, Jiageng and Wu, Xian and Zheng, Yefeng and Yang, Jie},
  journal={Journal of Biomedical Informatics},
  pages={104942},
  year={2025},
  publisher={Elsevier}
}

@article{liu2025improving,
  title={Improving large language model applications in biomedicine with retrieval-augmented generation: a systematic review, meta-analysis, and clinical development guidelines},
  author={Liu, Siru and McCoy, Allison B and Wright, Adam},
  journal={Journal of the American Medical Informatics Association},
  volume={32},
  number={4},
  pages={605--615},
  year={2025},
  publisher={Oxford University Press}
}

@inproceedings{vtyurina2018cues,
  author    = {Vtyurina, Alexandra and Fourney, Adam},
  title     = {Exploring the Role of Conversational Cues in Guided Task Support with Virtual Assistants},
  booktitle = {Proceedings of the 2018 CHI Conference on Human Factors in Computing Systems},
  series    = {CHI '18},
  year      = {2018},
  publisher = {Association for Computing Machinery},
  address   = {New York, NY, USA},
  articleno = {208},
  numpages  = {7},
  doi       = {10.1145/3173574.3173782}
}

@inproceedings{lee2024gazepointar,
  author    = {Lee, Jaewook and Wang, Jun and Brown, Elizabeth and Chu, Liam and Rodriguez, Sebastian S. and Froehlich, Jon E.},
  title     = {{GazePointAR}: A Context-Aware Multimodal Voice Assistant for Pronoun Disambiguation in Wearable Augmented Reality},
  booktitle = {Proceedings of the 2024 CHI Conference on Human Factors in Computing Systems},
  series    = {CHI '24},
  year      = {2024},
  publisher = {Association for Computing Machinery},
  address   = {New York, NY, USA},
  articleno = {408},
  numpages  = {20},
  doi       = {10.1145/3613904.3642230}
}

@article{arakawa2024prism,
  author    = {Arakawa, Riku and Lehman, Jill Fain and Goel, Mayank},
  title     = {{PrISM-Q\&A}: Step-Aware Voice Assistant on a Smartwatch Enabled by Multimodal Procedure Tracking and Large Language Models},
  journal   = {Proceedings of the ACM on Interactive, Mobile, Wearable and Ubiquitous Technologies},
  year      = {2024},
  volume    = {8},
  number    = {4},
  articleno = {180},
  numpages  = {26},
  month     = dec,
  doi       = {10.1145/3699759}
}

@inproceedings{huh2025vid2coach,
  author    = {Huh, Mina and Xue, Zihui and Das, Ujjaini and Ashutosh, Kumar and Grauman, Kristen and Pavel, Amy},
  title     = {{Vid2Coach}: Transforming How-To Videos into Task Assistants},
  booktitle = {Proceedings of the 38th Annual ACM Symposium on User Interface Software and Technology},
  series    = {UIST '25},
  year      = {2025},
  publisher = {Association for Computing Machinery},
  address   = {New York, NY, USA},
  articleno = {46},
  numpages  = {24},
  doi       = {10.1145/3746059.3747612}
}

@inproceedings{ning2025aroma,
  author    = {Ning, Zheng and Li, Leyang and Killough, Daniel and Seo, JooYoung and Carrington, Patrick and Tian, Yapeng and Zhao, Yuhang and Li, Franklin Mingzhe and Li, Toby Jia-Jun},
  title     = {{AROMA}: Mixed-Initiative {AI} Assistance for Non-Visual Cooking by Grounding Multimodal Information Between Reality and Videos},
  booktitle = {Proceedings of the 38th Annual ACM Symposium on User Interface Software and Technology},
  series    = {UIST '25},
  year      = {2025},
  publisher = {Association for Computing Machinery},
  address   = {New York, NY, USA},
  articleno = {144},
  numpages  = {15},
  doi       = {10.1145/3746059.3747650}
}

@inproceedings{arakawa2025scaling,
  author    = {Arakawa, Riku and Patidar, Prasoon and Page, Will and Lehman, Jill and Goel, Mayank},
  title     = {Scaling Context-Aware Task Assistants that Learn from Demonstration and Adapt through Mixed-Initiative Dialogue},
  booktitle = {Proceedings of the 38th Annual ACM Symposium on User Interface Software and Technology},
  series    = {UIST '25},
  year      = {2025},
  publisher = {Association for Computing Machinery},
  address   = {New York, NY, USA},
  numpages  = {19},
  doi       = {10.1145/3746059.3747700}
}

@inproceedings{Jaber2024Cooking,
  author    = {Jaber, Razan and Zhong, Sabrina and Kuoppam{\"a}ki, Sanna and Hosseini, Aida and Gessinger, Iona and Brumby, Duncan P. and Cowan, Benjamin R. and McMillan, Donald},
  title     = {Cooking With Agents: Designing Context-aware Voice Interaction},
  booktitle = {Proceedings of the 2024 CHI Conference on Human Factors in Computing Systems},
  series    = {CHI '24},
  year      = {2024},
  publisher = {Association for Computing Machinery},
  address   = {New York, NY, USA},
  articleno = {394},
  numpages  = {13},
  doi       = {10.1145/3613904.3642183}
}

@article{singhal2023medpalm,
  title = {Large language models encode clinical knowledge},
  author = {Singhal, Karan and Azizi, Shekoofeh and Tu, Tao and Mahdavi, S. Sara and Wei, Jason and Chung, Hyung Won and Scales, Nathan and Tanwani, Ajay and Cole-Lewis, Heather and Pfohl, Stephen and others},
  journal = {Nature},
  volume = {620},
  number = {7972},
  pages = {172--180},
  year = {2023},
  publisher = {Nature Publishing Group},
  doi = {10.1038/s41586-023-06291-2}
}

@article{zakka2024almanac,
  title = {Almanac --- Retrieval-Augmented Language Models for Clinical Medicine},
  author = {Zakka, Cyril and Shad, Rohan and Chaurasia, Akash and Dalal, Alex R. and Kim, Jennifer L. and Moor, Michael and Fong, Robyn and Phillips, Curran and Alexander, Kevin and Ashley, Euan and Boyd, Jack and Boyd, Kathleen and Hirsch, Karen and Langlotz, Curt and Lee, Rita and Melia, Joanna and Nelson, Joanna and Sallam, Karim and Tullis, Stacey and Vogelsong, Melissa Ann and Cunningham, John Patrick and Hiesinger, William},
  journal = {NEJM AI},
  volume = {1},
  number = {2},
  pages = {AIoa2300068},
  year = {2024},
  publisher = {Massachusetts Medical Society},
  doi = {10.1056/AIoa2300068}
}

@inproceedings{xiong2024medrag,
  title = {Benchmarking Retrieval-Augmented Generation for Medicine},
  author = {Xiong, Guangzhi and Jin, Qiao and Lu, Zhiyong and Zhang, Aidong},
  booktitle = {Findings of the Association for Computational Linguistics: ACL 2024},
  pages = {6233--6251},
  year = {2024},
  address = {Bangkok, Thailand},
  publisher = {Association for Computational Linguistics},
  doi = {10.18653/v1/2024.findings-acl.372}
}

@inproceedings{xiong2024confidence,
  title = {Can LLMs Express Their Uncertainty? An Empirical Evaluation of Confidence Elicitation in LLMs},
  author = {Xiong, Miao and Hu, Zhiyuan and Lu, Xinyang and Li, Yifei and Fu, Jie and He, Junxian and Hooi, Bryan},
  booktitle = {The Twelfth International Conference on Learning Representations (ICLR)},
  year = {2024}
}

@article{wen2025abstention,
  title = {Know Your Limits: A Survey of Abstention in Large Language Models},
  author = {Wen, Bingbing and Yao, Jihan and Feng, Shangbin and Xu, Chenjun and Tsvetkov, Yulia and Howe, Bill and Wang, Lucy Lu},
  journal = {Transactions of the Association for Computational Linguistics},
  volume = {13},
  pages = {529--556},
  year = {2025},
  publisher = {MIT Press},
  doi = {10.1162/tacl_a_00754}
}

@inproceedings{jo2023carecall,
  title = {Understanding the Benefits and Challenges of Deploying Conversational AI Leveraging Large Language Models for Public Health Intervention},
  author = {Jo, Eunkyung and Epstein, Daniel A. and Jung, Hyunhoon and Kim, Young-Ho},
  booktitle = {Proceedings of the 2023 CHI Conference on Human Factors in Computing Systems},
  series = {CHI '23},
  year = {2023},
  address = {Hamburg, Germany},
  publisher = {Association for Computing Machinery},
  doi = {10.1145/3544548.3581503}
}

@article{yang2024talk2care,
  title = {Talk2Care: An LLM-based Voice Assistant for Communication between Healthcare Providers and Older Adults},
  author = {Yang, Ziqi and Xu, Xuhai and Yao, Bingsheng and Rogers, Ethan and Zhang, Shao and Intille, Stephen and Shara, Nawar and Gao, Guodong Gordon and Wang, Dakuo},
  journal = {Proceedings of the ACM on Interactive, Mobile, Wearable and Ubiquitous Technologies},
  volume = {8},
  number = {2},
  pages = {73:1--73:35},
  year = {2024},
  publisher = {Association for Computing Machinery},
  doi = {10.1145/3659625}
}

@article{tu2025amie,
  title = {Towards conversational diagnostic artificial intelligence},
  author = {Tu, Tao and Schaekermann, Mike and Palepu, Anil and Saab, Khaled and Freyberg, Jan and Tanno, Ryutaro and Wang, Amy and Li, Brenna and Amin, Mohamed and Cheng, Yong and others},
  journal = {Nature},
  volume = {642},
  number = {8067},
  pages = {442--450},
  year = {2025},
  publisher = {Nature Publishing Group},
  doi = {10.1038/s41586-025-08866-7}
}

@article{laranjo2018conversational,
  title = {Conversational agents in healthcare: a systematic review},
  author = {Laranjo, Liliana and Dunn, Adam G and Tong, Huong Ly and Kocaballi, Ahmet Baki and Chen, Jessica and Bashir, Rabia and Surian, Didi and Gallego, Blanca and Magrabi, Farah and Lau, Annie Y S and Coiera, Enrico},
  journal = {Journal of the American Medical Informatics Association},
  volume = {25},
  number = {9},
  pages = {1248--1258},
  year = {2018},
  publisher = {Oxford University Press},
  doi = {10.1093/jamia/ocy072}
}

@article{bickmore2018safety,
  title = {Patient and Consumer Safety Risks When Using Conversational Assistants for Medical Information: An Observational Study of {Siri}, {Alexa}, and {Google Assistant}},
  author = {Bickmore, Timothy W and Trinh, Ha and Olafsson, Stefan and O'Leary, Teresa K and Asadi, Reza and Rickles, Nathaniel M and Cruz, Ricardo},
  journal = {Journal of Medical Internet Research},
  volume = {20},
  number = {9},
  pages = {e11510},
  year = {2018},
  publisher = {JMIR Publications},
  doi = {10.2196/11510}
}

@article{miner2016smartphone,
  title = {Smartphone-Based Conversational Agents and Responses to Questions About Mental Health, Interpersonal Violence, and Physical Health},
  author = {Miner, Adam S and Milstein, Arnold and Schueller, Stephen and Hegde, Roshini and Mangurian, Christina and Linos, Eleni},
  journal = {JAMA Internal Medicine},
  volume = {176},
  number = {5},
  pages = {619--625},
  year = {2016},
  publisher = {American Medical Association},
  doi = {10.1001/jamainternmed.2016.0400}
}

@article{ayers2023comparing,
  title = {Comparing Physician and Artificial Intelligence Chatbot Responses to Patient Questions Posted to a Public Social Media Forum},
  author = {Ayers, John W and Poliak, Adam and Dredze, Mark and Leas, Eric C and Zhu, Zechariah and Kelley, Jessica B and Faix, Dennis J and Goodman, Aaron M and Longhurst, Christopher A and Hogarth, Michael and Smith, Davey M},
  journal = {JAMA Internal Medicine},
  volume = {183},
  number = {6},
  pages = {589--596},
  year = {2023},
  publisher = {American Medical Association},
  doi = {10.1001/jamainternmed.2023.1838}
}

@article{fitzpatrick2017woebot,
  title = {Delivering Cognitive Behavior Therapy to Young Adults With Symptoms of Depression and Anxiety Using a Fully Automated Conversational Agent ({Woebot}): A Randomized Controlled Trial},
  author = {Fitzpatrick, Kathleen Kara and Darcy, Alison and Vierhile, Molly},
  journal = {JMIR Mental Health},
  volume = {4},
  number = {2},
  pages = {e19},
  year = {2017},
  publisher = {JMIR Publications},
  doi = {10.2196/mental.7785}
}

@article{pradhan2020use,
  title = {Use of Intelligent Voice Assistants by Older Adults with Low Technology Use},
  author = {Pradhan, Alisha and Lazar, Amanda and Findlater, Leah},
  journal = {ACM Transactions on Computer-Human Interaction},
  volume = {27},
  number = {4},
  articleno = {31},
  numpages = {27},
  year = {2020},
  publisher = {Association for Computing Machinery},
  address = {New York, NY, USA},
  doi = {10.1145/3373759}
}

@inproceedings{zubatiy2021empowering,
  title = {Empowering Dyads of Older Adults With Mild Cognitive Impairment And Their Care Partners Using Conversational Agents},
  author = {Zubatiy, Tamara and Vickers, Kayci L and Mathur, Niharika and Mynatt, Elizabeth D},
  booktitle = {Proceedings of the 2021 CHI Conference on Human Factors in Computing Systems},
  series = {CHI '21},
  year = {2021},
  publisher = {Association for Computing Machinery},
  address = {New York, NY, USA},
  articleno = {657},
  numpages = {13},
  doi = {10.1145/3411764.3445124}
}

@book{suchman1987plans,
  title = {Plans and Situated Actions: The Problem of Human-Machine Communication},
  author = {Suchman, Lucy A},
  year = {1987},
  publisher = {Cambridge University Press},
  address = {Cambridge, UK}
}

@incollection{clark1991grounding,
  title = {Grounding in Communication},
  author = {Clark, Herbert H and Brennan, Susan E},
  booktitle = {Perspectives on Socially Shared Cognition},
  editor = {Resnick, Lauren B and Levine, John M and Teasley, Stephanie D},
  pages = {127--149},
  year = {1991},
  publisher = {American Psychological Association},
  address = {Washington, DC, USA},
  doi = {10.1037/10096-006}
}

@article{clark1986referring,
  title = {Referring as a collaborative process},
  author = {Clark, Herbert H and Wilkes-Gibbs, Deanna},
  journal = {Cognition},
  volume = {22},
  number = {1},
  pages = {1--39},
  year = {1986},
  publisher = {Elsevier},
  doi = {10.1016/0010-0277(86)90010-7}
}

@inproceedings{horvitz1999mixed,
  title = {Principles of Mixed-Initiative User Interfaces},
  author = {Horvitz, Eric},
  booktitle = {Proceedings of the SIGCHI Conference on Human Factors in Computing Systems},
  series = {CHI '99},
  pages = {159--166},
  year = {1999},
  publisher = {Association for Computing Machinery},
  address = {New York, NY, USA},
  doi = {10.1145/302979.303030}
}

@inproceedings{porcheron2018voice,
  title = {Voice Interfaces in Everyday Life},
  author = {Porcheron, Martin and Fischer, Joel E and Reeves, Stuart and Sharples, Sarah},
  booktitle = {Proceedings of the 2018 CHI Conference on Human Factors in Computing Systems},
  series = {CHI '18},
  articleno = {640},
  numpages = {12},
  year = {2018},
  publisher = {Association for Computing Machinery},
  address = {New York, NY, USA},
  doi = {10.1145/3173574.3174214}
}

@inproceedings{myers2018patterns,
  title = {Patterns for How Users Overcome Obstacles in Voice User Interfaces},
  author = {Myers, Chelsea and Furqan, Anushay and Nebolsky, Jessica and Caro, Karina and Zhu, Jichen},
  booktitle = {Proceedings of the 2018 CHI Conference on Human Factors in Computing Systems},
  series = {CHI '18},
  articleno = {6},
  numpages = {7},
  year = {2018},
  publisher = {Association for Computing Machinery},
  address = {New York, NY, USA},
  doi = {10.1145/3173574.3173580}
}

@inproceedings{beneteau2019communication,
  title = {Communication Breakdowns Between Families and {Alexa}},
  author = {Beneteau, Erin and Richards, Olivia K and Zhang, Mingrui and Kientz, Julie A and Yip, Jason and Hiniker, Alexis},
  booktitle = {Proceedings of the 2019 CHI Conference on Human Factors in Computing Systems},
  series = {CHI '19},
  articleno = {243},
  numpages = {13},
  year = {2019},
  publisher = {Association for Computing Machinery},
  address = {New York, NY, USA},
  doi = {10.1145/3290605.3300473}
}

@article{lee2004trust,
  title = {Trust in Automation: Designing for Appropriate Reliance},
  author = {Lee, John D and See, Katrina A},
  journal = {Human Factors},
  volume = {46},
  number = {1},
  pages = {50--80},
  year = {2004},
  publisher = {SAGE Publications},
  doi = {10.1518/hfes.46.1.50_30392}
}

@article{bucinca2021trust,
  title = {To Trust or to Think: Cognitive Forcing Functions Can Reduce Overreliance on {AI} in {AI}-assisted Decision-making},
  author = {Bu{\c{c}}inca, Zana and Malaya, Maja Barbara and Gajos, Krzysztof Z},
  journal = {Proceedings of the ACM on Human-Computer Interaction},
  volume = {5},
  number = {CSCW1},
  articleno = {188},
  numpages = {21},
  year = {2021},
  publisher = {Association for Computing Machinery},
  address = {New York, NY, USA},
  doi = {10.1145/3449287}
}

@inproceedings{kay2016ish,
  title = {When (ish) is My Bus? User-centered Visualizations of Uncertainty in Everyday, Mobile Predictive Systems},
  author = {Kay, Matthew and Kola, Tara and Hullman, Jessica R and Munson, Sean A},
  booktitle = {Proceedings of the 2016 CHI Conference on Human Factors in Computing Systems},
  series = {CHI '16},
  pages = {5092--5103},
  year = {2016},
  publisher = {Association for Computing Machinery},
  address = {New York, NY, USA},
  doi = {10.1145/2858036.2858558}
}

@inproceedings{kocielnik2019imperfect,
  title = {Will You Accept an Imperfect {AI}? Exploring Designs for Adjusting End-user Expectations of {AI} Systems},
  author = {Kocielnik, Rafal and Amershi, Saleema and Bennett, Paul N},
  booktitle = {Proceedings of the 2019 CHI Conference on Human Factors in Computing Systems},
  series = {CHI '19},
  articleno = {411},
  numpages = {14},
  year = {2019},
  publisher = {Association for Computing Machinery},
  address = {New York, NY, USA},
  doi = {10.1145/3290605.3300641}
}

@inproceedings{yang2020reexamining,
  title = {Re-examining Whether, Why, and How Human-{AI} Interaction Is Uniquely Difficult to Design},
  author = {Yang, Qian and Steinfeld, Aaron and Ros{\'e}, Carolyn and Zimmerman, John},
  booktitle = {Proceedings of the 2020 CHI Conference on Human Factors in Computing Systems},
  series = {CHI '20},
  pages = {1--13},
  year = {2020},
  publisher = {Association for Computing Machinery},
  address = {New York, NY, USA},
  doi = {10.1145/3313831.3376301}
}

@article{lau2018alexa,
  title = {Alexa, Are You Listening? Privacy Perceptions, Concerns and Privacy-seeking Behaviors with Smart Speakers},
  author = {Lau, Josephine and Zimmerman, Benjamin and Schaub, Florian},
  journal = {Proceedings of the ACM on Human-Computer Interaction},
  volume = {2},
  number = {CSCW},
  articleno = {102},
  numpages = {31},
  year = {2018},
  publisher = {Association for Computing Machinery},
  address = {New York, NY, USA},
  doi = {10.1145/3274371}
}

@inproceedings{dahlback1993wizard,
  title = {Wizard of {Oz} Studies: Why and How},
  author = {Dahlb{\"a}ck, Nils and J{\"o}nsson, Arne and Ahrenberg, Lars},
  booktitle = {Proceedings of the 1st International Conference on Intelligent User Interfaces},
  series = {IUI '93},
  pages = {193--200},
  year = {1993},
  publisher = {Association for Computing Machinery},
  address = {New York, NY, USA},
  doi = {10.1145/169891.169968}
}

@inproceedings{zheng2023judging,
  title = {Judging {LLM}-as-a-Judge with {MT-Bench} and {Chatbot Arena}},
  author = {Zheng, Lianmin and Chiang, Wei-Lin and Sheng, Ying and Zhuang, Siyuan and Wu, Zhanghao and Zhuang, Yonghao and Lin, Zi and Li, Zhuohan and Li, Dacheng and Xing, Eric P and Zhang, Hao and Gonzalez, Joseph E and Stoica, Ion},
  booktitle = {Advances in Neural Information Processing Systems},
  volume = {36},
  series = {NeurIPS '23},
  year = {2023},
  pages = {46595--46623}
}


\FloatBarrier
\clearpage
\appendix

\section{Additional LLM Evaluation Examples}
\label{app:llm-examples}

\subsection{Errors Persisting Under Explicit Guardrails}
\label{app:guardrail-examples}

Table~\ref{tab:guardrail-examples} provides examples showing that each of
the five error categories observed in our primary evaluation persisted when
models were explicitly instructed to avoid these behaviors.

\subsection{Premature Advancement After Adding Procedural Context}
\label{app:premature}

\begin{table*}[t]
\caption{Examples showing that all five error categories persisted under explicit
guardrails. Error categories are defined in Table~\ref{tab:error-categories}.
Each model response was produced with the guardrail addendum appended to the prompt.}
\label{tab:guardrail-examples}
\Description{
The table gives one example demonstrating that each of the five error
categories persists even when explicit error guardrails are included. The
Wrong example recommends soap and water instead of clean hands for washing the
mock wound. The Visual example confirms an unseen Vaseline packet. The
Premature example advances from drying the hands to placing the supply bag on
the counter. The Unanswered example instructs the patient to remove the Telfa
dressing without answering how to identify it. The Ignored example treats the
counter's cleanliness as unknown despite the dialogue already establishing
that the patient cleaned it. Each example contrasts the guarded model response
with the expert reference response.
}
\small
\setlength{\tabcolsep}{6pt}
\renewcommand{\arraystretch}{1.08}

\begin{tabularx}{\textwidth}{
    @{}
    p{0.12\textwidth}|
    X
    @{}
}
\toprule
\textbf{Error} &
\textbf{Example under explicit guardrail} \\
\midrule

\textbf{Wrong} &
Context: The band-aid on the patient's skin is the fake wound. The patient was asked to wash it and checked whether it should come off first.
Patient: ``Should I remove the bandage? Okay, gently wash the… fake wound with soap and water.''
GPT-4.1: ``Yes, please remove the bandage before gently washing the fake wound with soap and water. According to the wound care instructions, you should remove the bandage before cleaning the wound.''
Expert: ``Do not remove the band-aid that is the fake wound, just wash it gently with soap and water.''\\

\midrule

\textbf{Visual} &
\textbf{Context:} The voice-only assistant had no visual access while the patient
was trying to identify the Vaseline packet. \newline
\textbf{Patient:} ``This is Vaseline, right?'' \newline
\textbf{Claude Haiku 4.5:} ``\textbf{Yes, that appears to be the Vaseline.}
Please go ahead and open it so you can apply it to your wound with the Q-tips.'' \newline
\textbf{Expert:} ``The Vaseline is in a labeled packet.'' \\

\midrule

\textbf{Premature} &
\textbf{Context:} The patient had finished washing their hands and was still
drying them; placing the supply bag on the counter was the subsequent step. \newline
\textbf{Patient:} ``Washing my hands with soap and water, done \ldots\ dry 'em.'' \newline
\textbf{GPT-4o-mini:} ``Great job! \textbf{Now, please place the bag of supplies
on the clean counter.}'' \newline
\textbf{Expert:} ``Yes, please dry your hands. You can use a clean paper towel.'' \\

\midrule

\textbf{Unanswered} &
\textbf{Context:} The patient was trying to identify the Telfa non-stick dressing. \newline
\textbf{Patient:} ``What is it you want, the non-stick?'' \newline
\textbf{GPT-4o:} ``Please remove the Telfa non-stick dressing from the supply bag.'' \newline
\textbf{Expert:} ``The Telfa non-stick dressing is in a packet with the word
Telfa written on it.'' \\

\midrule

\textbf{Ignored} &
\textbf{Context:} The patient had already cleaned the counter using the cleaning
spray and a paper towel. \newline
\textbf{Patient:} ``How do you know it's clean?'' \newline
\textbf{GPT-4o:} ``\textbf{I don't know if the counter is clean.} Please ensure
the counter is clean by wiping it down with a disinfectant before placing the bag
of supplies on it.'' \newline
\textbf{Expert:} ``The counter should be clean since you have cleaned it with the
cleaning spray and a paper towel.'' \\

\bottomrule
\end{tabularx}
\end{table*}

\begin{table*}[t]
\caption{Examples of premature advancement introduced after providing the complete procedure, with one example from each of the six models for which premature responses increased. 
Responses are shortened for space; bold text indicates the newly introduced future action.}
\label{tab:premature-examples}
\Description{
The table presents paired examples for the six models whose premature
advancement increased after the complete procedure was added. Each row shows
the model and procedure step, the participant's request, the response with
interaction context alone, and the response after adding the procedure. In
each example, the interaction-only response appropriately addresses the
current step, whereas the procedure-conditioned response additionally
introduces a future action before completion of the current action has been
established. The six examples span Claude 3.5 Sonnet, Gemini 1.5 Flash,
Gemini 1.5 Pro, GPT-4.1, GPT-4o, and GPT-4o-mini.
}
\small
\begin{tabularx}{\textwidth}{@{}p{0.15\textwidth}|p{0.19\textwidth}|X|X@{}}
\toprule
\textbf{Model / Step} &
\textbf{Participant request} &
\textbf{Interaction only} &
\textbf{Interaction + Procedure} \\
\midrule

Claude 3.5 Sonnet / S4 &
\textit{``Oh okay, and throw it away I'm assuming.''} &
``Yes, please remove the gauze and dispose of it in the trash.'' 
\newline \textit{Acceptable} &
``Yes, please throw away the used gauze in the trash... 
\textbf{You can proceed to the next step of gently washing the fake wound with soap and water.}''
\newline \textit{Premature} \\

\midrule

Gemini 1.5 Flash / S1 &
\textit{``Do I actually do it?''} &
``Yes, please clean the counter with the paper towel and spray bottle... Let me know when you're finished.''
\newline \textit{Acceptable} &
``Yes, please clean the counter... Let me know when you're finished... 
\textbf{Then we can wash your hands.}''
\newline \textit{Premature} \\

\midrule

Gemini 1.5 Pro / S2 &
\textit{``So I have to get up to the...''} &
``Yes, please wash your hands at the sink with soap and water.''
\newline \textit{Acceptable} &
``Yes, please wash your hands with soap and water at the sink. 
\textbf{Afterward, please return and place the bag of supplies on the clean counter.}''
\newline \textit{Premature} \\

\midrule

GPT-4.1 / S3 &
\textit{``...Tefla, is it Tefla or Telfa?''} &
``It's actually called `Telfa,' spelled T-E-L-F-A...''
\newline \textit{Acceptable} &
``It's actually called `Telfa,' spelled T-E-L-F-A... 
\textbf{Please continue by removing the gauze covering your fake wound.}''
\newline \textit{Premature} \\

\midrule

GPT-4o / S2 &
\textit{``Do I do it?''} &
``Yes, please wash your hands thoroughly with soap and water...''
\newline \textit{Acceptable} &
``Yes, please wash your hands thoroughly with soap and water... 
\textbf{Once your hands are clean, you can proceed to the next step, which is placing the bag of supplies on the clean counter.}''
\newline \textit{Premature} \\

\midrule

GPT-4o-mini / S2 &
\textit{``And that's for real?''} &
``Yes, please wash your hands with soap and water for real...''
\newline \textit{Acceptable} &
``Yes, please wash your hands with soap and water for real... 
\textbf{After washing, you can place the bag of supplies on the clean counter.}''
\newline \textit{Premature} \\

\bottomrule
\end{tabularx}
\end{table*}

Table~\ref{tab:premature-examples} provides paired examples of premature
advancement introduced after the complete procedure was added. We include one
example from each of the six models for which this pattern increased.

\end{document}